\documentclass[12pt,preprint]{emulateapj}

\submitted{Accepted for publication in ApJ 2009 February 5}
\shorttitle{2MASS 06164006$-$6407194}
\shortauthors{Cushing et al.}

\newcommand\teff{\mbox{$T_\mathrm{eff}$}}

\newcommand\obj{\mbox{2MASS J0616$-$6407}}
\newcommand\sdone{\mbox{2MASS J0532$+$8246}}
\newcommand\sdtwo{\mbox{2MASS J1626$+$3925}}
\newcommand\sdthree{\mbox{SDSS J1256$-$0224}}

\begin{document}


\title{2MASS J06164006$-$6407194: The First Outer Halo L Subdwarf}


\author{Michael C. Cushing \& Dagny Looper}
\affil{Institute for Astronomy, University of Hawai'i, 2680 Woodlawn Drive, Honolulu, HI 96822}
\email{mcushing@ifa.hawaii.edu}

\author{Adam J. Burgasser}
\affil{Massachusetts Institute of Technology, Kavli Institute for Astrophysics and Space Research, Building 37, Room 664B, 77 Massachusetts Avenue, Cambridge, MA 02139}

\author{J. Davy Kirkpatrick}
\affil{Infrared Processing and Analysis Center, M/S 100-22, California Institute of Technology, Pasadena, CA 91225}

\author{Jacqueline Faherty}
\affil{Department of Physics and Astronomy, Stony Brook University, Stony Brook, NY, 11794}

\author{Kelle L. Cruz \& Anne Sweet}
\affil{Dept. of Astronomy, Caltech, MC 105-24, Pasadena, CA 91125}

\author{Robyn E. Sanderson} 
\affil{Massachusetts Institute of Technology, Kavli Institute for
  Astrophysics and Space Research, Building 37, Room 664B, 77
  Massachusetts Avenue, Cambridge, MA 02139}

\begin{abstract}

  We present the serendipitous discovery of an L subdwarf, 2MASS
  J06164006$-$6407194, in a search of the Two Micron All Sky Survey for
  T dwarfs.  Its spectrum exhibits features indicative of both a cool
  and metal poor atmosphere including a heavily pressured-broadened
  \ion{K}{1} resonant doublet, \ion{Cs}{1} and \ion{Rb}{1} lines,
  molecular bands of CaH, TiO, CrH, FeH, and H$_2$O, and enhanced
  collision induced absorption of H$_2$.  We assign \obj\ a spectral
  type of sdL5 based on a comparison of its red optical spectrum to that
  of near solar-metallicity L dwarfs.  Its high proper motion ($\mu$ =
  1$\farcs$405$\pm$0.008 yr$^{-1}$), large radial velocity
  ($V_{\mathrm{rad}}$ = 454$\pm$15 km s$^{-1}$), estimated $uvw$
  velocities (94, $-$573, 125) km s$^{-1}$ and Galactic orbit with an
  apogalacticon at $\sim$29 kpc are indicative of membership in the
  outer halo making \obj\ the first ultracool member of this population.

\end{abstract}

\section{Introduction} \label{sec:Introduction}

Ultracool subdwarfs are metal-poor stars and brown dwarfs that have
spectral types later than $\sim$sdM7/esdM7 \citep{2005ESASP.560..237B}
and \teff\ $\la$ 3000 K \citep{2000ApJ...535..965L}.  Since cool
subdwarfs are typically members of the thick disk or halo population,
they exhibit large space motions relative to the Sun \citep[$<V>$=$-$202
km s$^{-1}$; ][]{1997AJ....113..806G} and thus are often identified as
high proper-motion stars \citep[e.g.,
][]{1971lpms.book.....G,1978LowOB...8...89G,1979lccs.book.....L,2004A&A...421..763P,2005AJ....129.1483L}.
Due to their intrinsic faintness and paucity in the solar neighborhood,
only $\sim$40 ultracool subdwarfs are currently known
\citep{2007ApJ...657..494B,2008arXiv0804.1731L}.  Such small numbers
stand in stark contrast to the over one thousand near solar-metallicity
ultracool dwarfs (spectral types later than M7 and \teff\ $\la$ 2400 K)
that have been discovered in the Two Micron All Sky Survey \citep[2MASS;
][]{2006AJ....131.1163S}, the Sloan Digital Sky Survey \citep[SDSS;
][]{2000AJ....120.1579Y}, and the Deep Near Infrared Southern Sky Survey
\citep{1997Msngr..87...27E}, many of which populate the L and T spectral
classes\footnote{A compendium of known L and T dwarfs is given at
  http://DwarfArchives.org.} \citep{2005ARA&A..43..195K}.

Of the roughly 40 ultracool subdwarfs known, only three, 2MASS
J0532346$+$8246465 \citep[hereafter \sdone; ][]{2003ApJ...592.1186B},
2MASS J16262034$+$3925190 \citep[hereafter \sdtwo;
][]{2004ApJ...614L..73B}, and SDSS 1256$-$0224 \citep{sivarani04} are
classified as L subdwarfs\footnote{LSR 1610$-$0040 was originally
  classified as an L subdwarf by \citet{2003ApJ...591L..49L} but
  observations by \citet{2006AJ....131.1797C},
  \citet{2006AJ....131.1806R}, and \citet{2008arXiv0806.2336D} indicate
  that it is probably a peculiar M dwarf/subdwarf binary.}.  The spectra
of all three L subdwarfs exhibit \ion{Cs}{1} and \ion{Rb}{1} lines,
pressure-broadened \ion{K}{1} lines, CaH, FeH, CrH and H$_2$O bands, and
enhanced collision induced H$_2$ absorption \citep[CIA,
e.g.,][]{1969ApJ...156..989L,1997A&A...324..185B}, all features
indicative of very cool
\citep{1999ApJ...519..802K,2001ApJ...548..908L,2001AJ....121.1710R} and
metal-depleted atmospheres
\citep{1982PASAu...4..417B,1994ApJ...424..333S}.  In addition, they have
large proper motions (0$\farcs$6 to 2$\farcs$6 yr$^{-1}$) and radial
velocities ($-$130 to $-$260 km s$^{-1}$) consistent with halo
membership.  \citet{2008ApJ...672.1159B} recently measured the parallax
of \sdone\ and confirmed, based on its Galactic orbit, that it is indeed
a member of the (inner) halo.  Although no spectral classification
scheme exists for L subdwarfs, \citet{2007ApJ...657..494B,burgasser09a}
have tentatively assigned spectral types of sdL3.5, sdL4, and sdL7 to
\sdthree, \sdtwo, and \sdone, respectively, based on a comparison with
the red optical spectra of the near solar metallicity L dwarf standards
\citep{1999ApJ...519..802K}.

In this paper, we present the discovery of a fourth L subdwarf, 2MASS
J06164006$-$6407194 (hereafter \obj) found in a search of the 2MASS
database for T dwarfs.  The discovery of \obj, along with followup
near-infrared imaging and spectroscopy, and red-optical spectroscopy is
presented in \S2.  In \S3 we discuss the spectral classification and
kinematics of \obj.

\section{Observations}

\subsection{Source Identification}

\obj\ was identified as a T dwarf candidate by
\citet{2007AJ....134.1162L} in the course of a search for T dwarfs in
the 2MASS All Sky Survey Database.  It was selected based on its very
blue near-infrared color ($J-K_s$ $<$ 0.02) and lack of any optical
counterpart within 5$''$ in the USNO-A2.0 catalog or the Digital Sky
Survey (DSS) I and II images.  A finder chart for \obj\ is given in
Figure 1 and pertinent properties are listed in Table \ref{tab:ObsProp}.
Below we describe followup near-infrared imaging and spectroscopy and
red-optical spectroscopy.

\begin{figure} 
\includegraphics[width=3.5in]{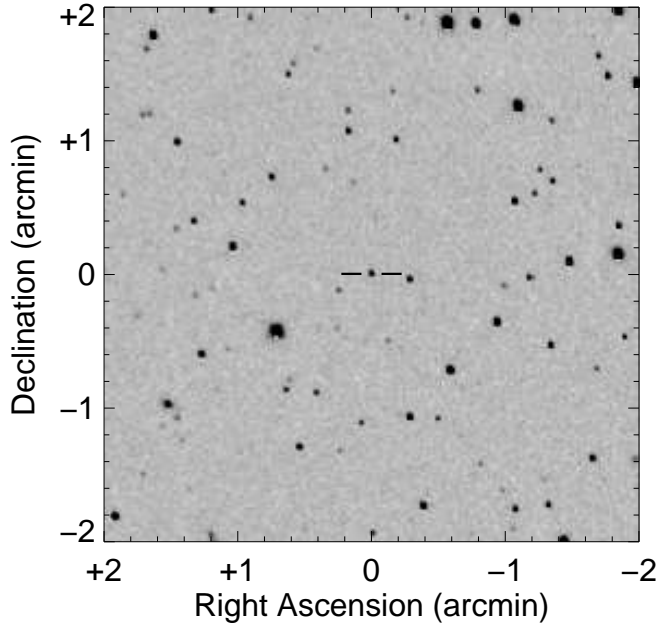}
\caption{\label{fig:Finder}$J$-band image of the field centered around
  \obj\ obtained with the CPAPIR on the 1.5-m CTIO on 2008-01-17 (UT).
  The field is 4$'$ on a side and North is up and East is to the left.}
\end{figure}

\subsection{Near Infrared Spectroscopy}

\obj\ was observed with the Ohio State InfraRed Imager/Spectrometer
(OSIRIS) mounted at the 4.1 m SOAR Telescope on 2007 March 12 (UT).  We
used the 1$\farcs$0 wide slit which provides a resolving power of
$R\equiv\lambda/\Delta \lambda \sim$ 1400 across the 1 to 2.2 $\mu$m
wavelength range in three spectral orders.  A series of six 180 sec
integrations were obtained along the 24$''$-long slit.  The A0 V star HD
53191 was also observed to correct for telluric absorption and to
flux-calibrate the final spectrum.  The data were reduced using a
modified version (v3.4) of the Spextool data reduction package
\citep{2004PASP..116..362C}.  Wavelength calibration was accomplished
using the OH airglow lines in the science frames.  The spectra from the
3 orders were then corrected for telluric absorption using the
observations of the A0 V standard stars and the technique described in
\citet{2003PASP..115..389V}.  The 3 spectral orders were then merged
into a single spectrum covering the 1.2$-$2.3 $\mu$m wavelength range.
The final spectrum was then rebinned to one pixel per resolution element
in order to increase the signal-to-noise ratio (S/N) of the spectrum.

\begin{figure} 
\includegraphics[width=3.5in]{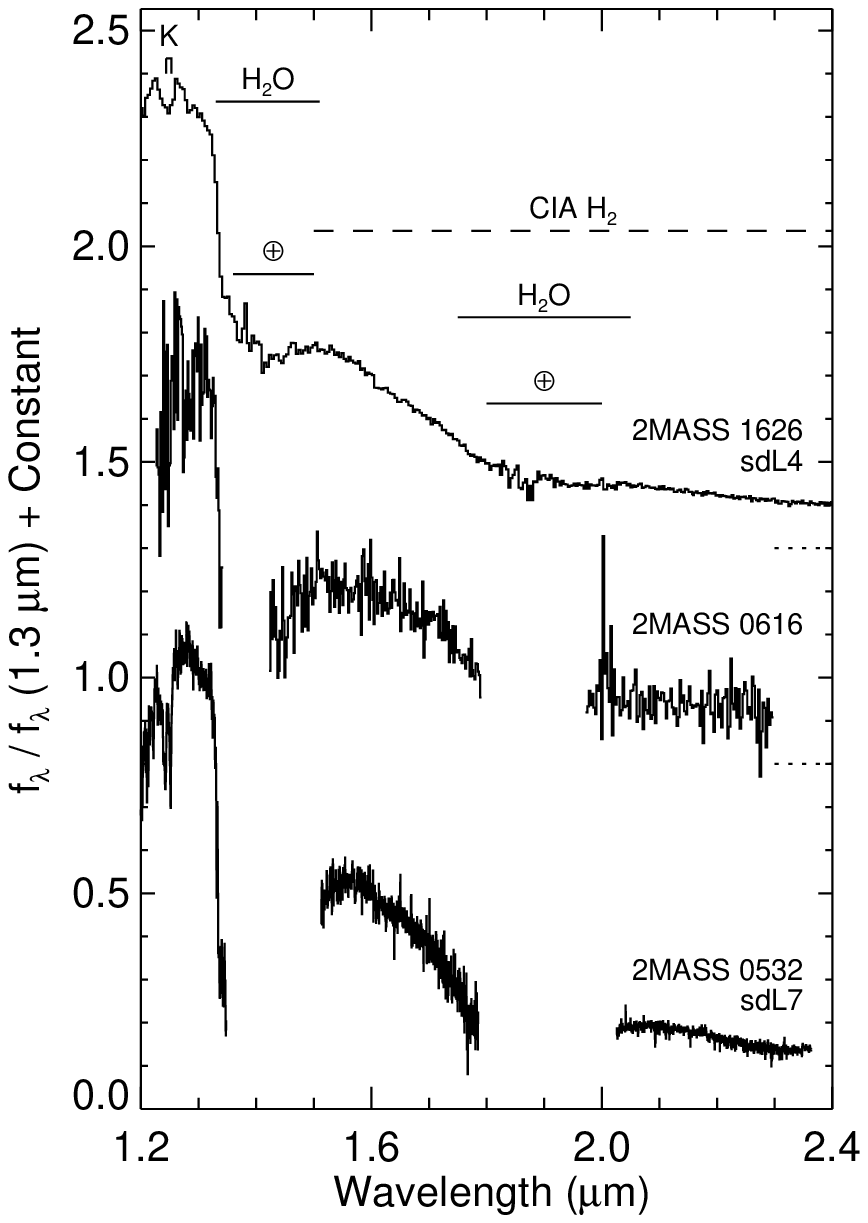}
\caption{\label{fig:NIRComp}Near infrared spectra of \sdtwo\,
  \citep[sdL4;][]{2007ApJ...657..494B}, \obj, and \sdone\,
  \citep[sdL7;][]{2003ApJ...592.1186B}.  The spectra are normalized to
  unity at 1.3 $\mu$m and offset by constants (\textit{dotted lines}).
  Prominent atomic and molecular absorption features are indicated.  The
  S/N of the \obj\ spectrum is not high enough to detect the 1.25 $\mu$m
  \ion{K}{1} doublet.}
\end{figure}

The spectrum of \obj\ is shown in Figure \ref{fig:NIRComp}
(\textit{middle}).  The S/N of the spectrum is low, ranging from $\sim$8
in the $J$ and $H$ bands to $\sim$4 in the $K$ band, but is nevertheless
adequate to confirm that \obj\ is not a T dwarf (the $J-K_s$ color of
\obj\ corresponds to a spectral type of $\sim$T4) since the spectrum
lacks the prominent CH$_4$ bands found in the spectra of T dwarfs
\citep{2002ApJ...564..421B,2002ApJ...564..466G}.  Also shown in Figure
\ref{fig:NIRComp} are the spectra of the two other L subdwarfs with
published spectra, \sdone\ \citep{2003ApJ...592.1186B} and \sdtwo\
\citep{2004ApJ...614L..73B}.  All three spectra exhibit blue continua
due to enhanced collision induced absorption (CIA) of H$_2$ centered at
2.4 $\mu$m and broad absorption bands of H$_2$O centered at 1.4 and 1.9
$\mu$m indicating that \obj\ is an L subdwarf.

\begin{deluxetable}{ll}
\tablecolumns{2}
\tabletypesize{\scriptsize} 
\tablewidth{0pc}
\tablecaption{\label{tab:ObsProp}Properties of 2MASS J0616$-$6407}
\tablehead{
\colhead{Parameter} & 
\colhead{Value}} 
 
\startdata
 
\multicolumn{2}{c}{Observed} \\
\hline

$\alpha$(J2000.0)\tablenotemark{a}  & 6$^h$16$^m$40$\fs$06 \\
$\delta$(J2000.0)\tablenotemark{a}  & $-$64$^\circ$07$'$19$\farcs$4 \\
2MASS $J$                           & 16.403$\pm$0.113 mag \\
2MASS $H$                           & 16.275$\pm$0.228 mag \\
2MASS $K_s$                         & $>$16.381\tablenotemark{b} mag \\
$\mu_\alpha$ (J2000.0)              & 1405$\pm$8 mas yr$^{-1}$\\ 
$\mu_\delta$ (J2000.0)              & $-$51$\pm$18 mas yr$^{-1}$ \\
$V_{\mathrm{rad}}$                  & 454$\pm$15 km s$^{-1}$ \\

\hline
\multicolumn{2}{c}{Estimated} \\
\hline
$d$                                 &  57$\pm$9 pc \\
$V_{\mathrm{tan}}$                  & 379$\pm$60 km s$^{-1}$ \\
$u$\tablenotemark{c}                & 94$\pm$10 km s$^{-1}$\\
$v$                                 & $-$573$\pm$31 km s$^{-1}$ \\
$w$                                 & 125$\pm$53 km s$^{-1}$ \\

\enddata

\tablenotetext{a}{2MASS coordinates at epoch 1998.95 (UT).}
\tablenotetext{b}{The read flag is 0 and the quality flag is U
  indicating that no flux was detected at the position and that the
  value is an upper limit.}  \tablenotetext{c}{We follow the convention
  that the $u$ is positive towards the Galactic Center ($l$=0, $b$=0).}

\end{deluxetable}

\subsection{Red Optical Spectroscopy}

\subsubsection{Magellan/LDSS-3}\label{sec:LDSS3}

Red optical spectra of \obj\ were obtained on 2006 May 10 (UT) with
LDSS-3 on Magellan.  The VPH-red grism (660 lines/mm) with a 0$\farcs$75
wide (4 pixels) longslit mask was used, with the slit aligned to the
parallactic angle.  This configuration provides 6050--10500 {\AA}
spectra across the entire chip with an average resolving power of $R$
$\approx$ 1800 and dispersion along the chip of $\sim$1.2~{\AA}/pixel.
The OG590 longpass filter was used to eliminate second order light
shortward of 6000~{\AA}.  Two slow-read exposures of 1800~s each were
obtained at an airmass of 1.49--1.60. We also observed the G2~V star HD
44611 immediately after and at a roughly similar airmass (1.77) for
telluric absorption correction.  The flux standard LTT 7987
\citep[a.k.a.\ GJ 2147; ][]{1994PASP..106..566H} was observed on the
previous night using an identical slit and grism combination. All
spectral observations were accompanied by HeNeAr arc lamp and flat-field
quartz lamp exposures for dispersion and pixel response calibration.

LDSS-3 data were reduced in the IRAF\footnote{IRAF is distributed by the
  National Optical Astronomy Observatories, which are operated by the
  Association of Universities for Research in Astronomy, Inc., under
  cooperative agreement with the National Science Foundation.}
environment \citep{1986SPIE..627..733T}.  Raw images were first
corrected for amplifier bias voltage, stitched together, and subtracted
by a median-combined set of slow-read bias frames taken during the
afternoon.  These processed images were then divided by a
median-combined, bias-subtracted and normalized set of flat field
frames.  The LTT 7987 and HD 44611 spectra were optimally extracted
first using the APALL task with background subtraction.  The spectrum of
\obj\ was then extracted using the G star dispersion trace as a
template.  Dispersion solutions were determined from arc lamp spectra
extracted using the same dispersion trace; solutions were accurate to
$\sim$0.08 pixels, or $\sim$0.1~{\AA}.  Flux calibration was determined
using the tasks STANDARD and SENSFUNC with observations of LTT 7987,
adequate over the spectral range 6000--10000 {\AA}.  Corrections to
telluric O$_2$ (6855--6955 {\AA} B-band, 7580--7740 {\AA} A-band) and
H$_2$O (7160--7340 {\AA}, 8125--8350~{\AA}, 9270--9680 {\AA}) absorption
bands were determined by linearly interpolating over these features in
the G dwarf spectrum, dividing by the uncorrected spectrum, and
multiplying the result with the spectrum of \obj.  The two spectra of
\obj\ were then coadded to improve S/N.  We measured the radial velocity
of \obj\ using the \ion{Rb}{1} (7948 \AA), \ion{Na}{1} (8183, 8195 \AA),
and \ion{Cs}{1} (8521 \AA) lines.  The observed line positions
determined by Gaussian fits were compared against the rest wavelengths
\citep{NIST} to derive a heliocentric radial velocity (after correcting
for the Earth's orbital velocity) of 480$\pm$30 km s$^{-1}$, where the
error in the radial velocities arises primarily from the scatter in the
line positions.  We defer further discussion of the spectrum to the next
section.

\subsubsection{Gemini/GMOS Spectroscopy}\label{sec:GMOS}

\obj\ was also observed with the Gemini Multi-Object Spectrograph
\citep{2004PASP..116..425H} mounted on the Gemini South Telescope on
2007 September 13 (UT).  We used the 0$\farcs$75 wide slit and R400
grating which yields a spectrum from 5900 to 10100 \AA\, with a
resolving power of $R\sim$ 1300.  The OG515 filter was used to suppress
stray light from shorter wavelengths.  A total of five 1800 sec
exposures were acquired at two different central wavelengths and at two
positions along the slit.  Standard calibration frames including
flat-field and CuAr lamp images as well as observations of the G2 V star
HD 60402 for telluric correction were also obtained.  The flux standard
EG 21 \citep{1994PASP..106..566H} was observed on 2007 September 14 (UT)
with the same instrument setup.

\begin{figure*}[t] 
\hspace{0.0in}
\centerline{\includegraphics[width=6in]{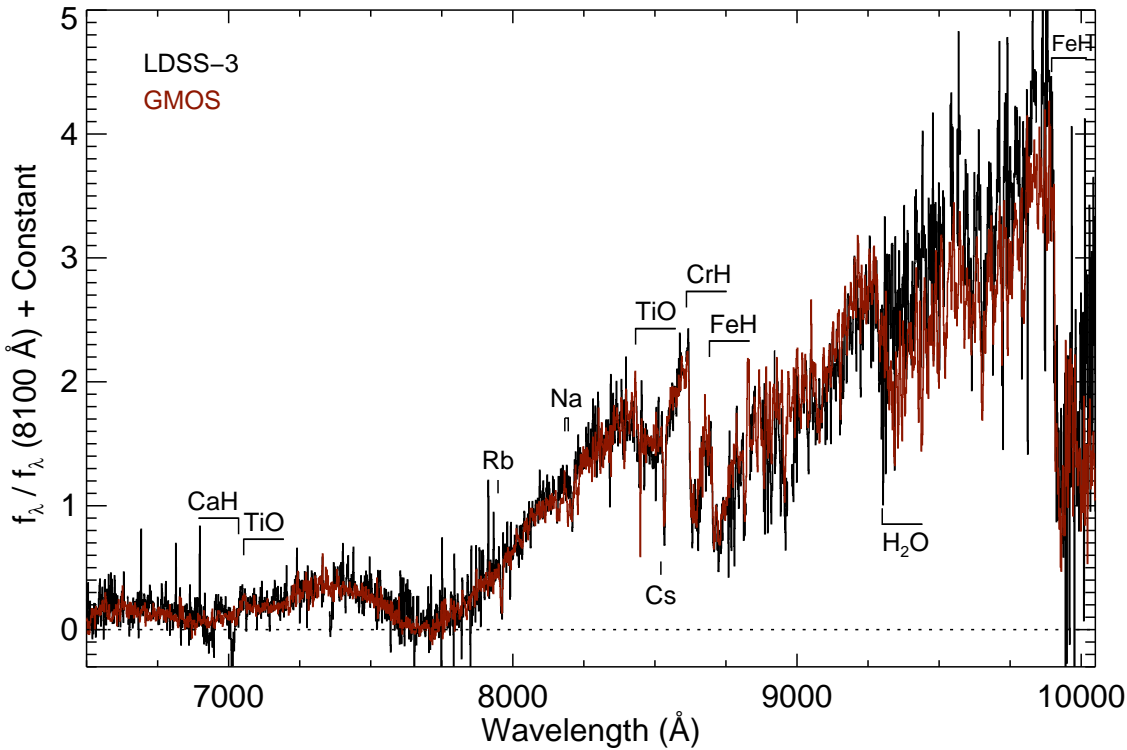}}
\caption{\label{fig:SpecComp}Red optical spectra of \obj\ obtained with
  LDSS-3/Magellan (black) and GMOS/Gemini (red).  The spectra are
  normalized to unity at 8100 \AA.\, Prominent atomic and molecular
  absorption features are indicated.}
\end{figure*}

The data were first reduced using the IRAF Gemini GMOS package.  The raw
frames were first bias subtracted (with a bias created using GBIAS) and
flat-fielded (with a normalized flat field created with GSFLAT).  The OH
sky lines were then subtracted and then spectra were optimally extracted
using the GSEXTRACT routine.  Wavelength calibration was achieved using
the CuAr lamp images.  The spectra were then flux calibrated using the
observations of EG 21 and the GSCALIBRATE routine.  Telluric correction
was achieved by linearly interpolating over regions of O$_2$ and H$_2$O
absorption in the spectrum of HD 60402, dividing by the raw spectrum of
HD 60402, and then multiplying the results into the flux-calibrated
spectrum of \obj.

Unfortunately, the resulting spectrum exhibits strong noise spikes at
the wavelengths of nearly all the OH sky lines.  Fringing in the
detector makes the subtraction of the OH sky lines using low order
polynomials difficult.  We therefore re-extracted the raw spectrum of
\obj\ so that we could first pair subtract the images with the \obj\ at
different positions along the slit.  Although this did not completely
eliminate the strong residuals, it did improve the spectrum
dramatically.  Finally we measured a radial velocity of $+$445 $\pm$18
km s$^{-1}$ using the same technique described in the previous section.
The radial velocity measurements from the LDSS-3 and GMOS spectra agree
within the errors and a weighted average of the two values gives
454$\pm$15 km s$^{-1}$.

The LDSS-3 and GMOS spectra are shown in Figure \ref{fig:SpecComp}.  The
GMOS spectrum has a S/N of 10 to 20 while the LDSS-3 has a lower S/N.
The spectra exhibit band heads of CaH (7035 \AA), TiO (7053, 8432 \AA),
CrH (8611 \AA), FeH (8692, 9896 \AA), and atomic lines of \ion{K}{1}
(7665, 7699 \AA), \ion{Rb}{1} (7800, 7948 \AA), \ion{Na}{1} (8183, 8195
\AA), and \ion{Cs}{1} (8521 \AA).  The heavily pressure-broadened
\ion{K}{1} doublet, strong CrH and FeH band heads, and \ion{Cs}{1} and
\ion{Rb}{1} lines are hallmark spectral features of L dwarfs
\citep{1999ApJ...519..802K} while the presence of the CaH and TiO band
heads are consistent with known late-type M and L subdwarf spectra
\citep{2007ApJ...657..494B}.  Overall the spectra agree well except
longward of $\sim$9200 \AA\ where the LDSS-3 spectrum is systematically
higher than the GMOS spectrum.  \citet{2007ApJ...657..494B} found a
similar offset between LDSS-3 and GMOS spectra of ultracool M subdwarfs
and ascribed it to an unknown flux calibration error and/or the fact
that the 9200 \AA\ telluric H$_2$O band was not corrected in their GMOS
data.  Our GMOS spectrum of \obj\ has been corrected for telluric
absorption so we can eliminate this as the cause of the mismatch.  Since
the cause remains unknown, we restrict further analysis to $\lambda <$
9200 \AA.

\begin{figure} 
\includegraphics[width=3.2in]{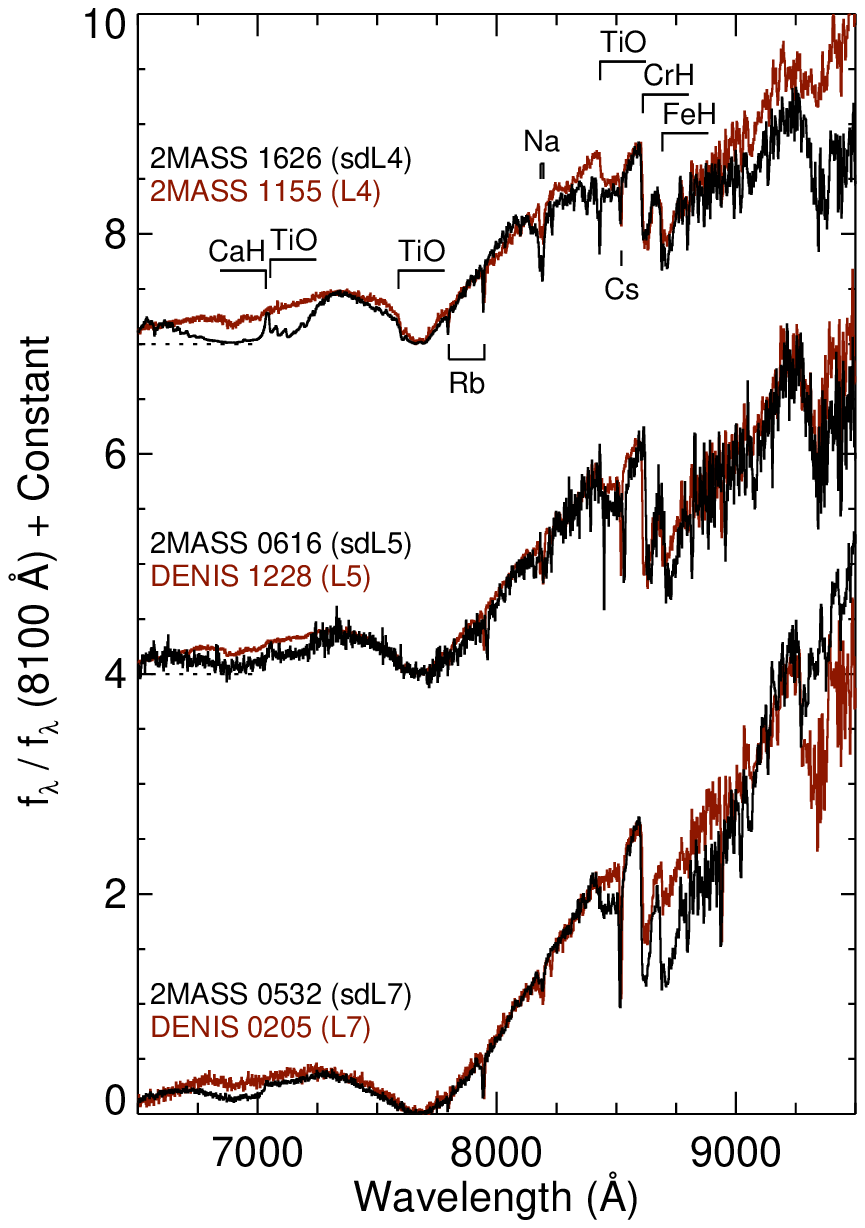}
\caption{\label{fig:OptComp}Red optical spectra of \sdtwo\,
  \citep[sdL4;][]{2007ApJ...657..494B}, \obj, and \sdone\,
  \citep[sdL7;][]{2003ApJ...592.1186B} compared to those of 2MASS
  1155009$+$230706 \citep[L4;][]{1999ApJ...519..802K}, DENIS-P
  J1228.2-1547 \citep[L5;][]{1999ApJ...519..802K} and DENIS-P
  J0205.4-1159 \citep[L7;][]{1999ApJ...519..802K}, respectively.  The
  spectra are normalized to unity and offset by constants
  (\textit{dotted lines)}.  Prominent atomic and molecular absorption
  features are indicated.}
\end{figure}

\subsection{Near Infrared Imaging}

An image of \obj\ was taken on 2008 Jan 17 (UT) using the 1.5m telescope
at CTIO with the Cam\'{e}ra PAnoramique Proche InfraRouge \citep[CPAPIR;
][]{2004SPIE.5492.1479A} wide field-IR imager in the $J$ band.  The
object was observed close to the meridian at an airmass of 1.2 and with
seeing conditions $\sim$1$\farcs$5.  We used a 9-point dither pattern
with 10$''$ offsets.  Each exposure was $\sim$60 seconds amounting to a
total integration time of $\sim$9 minutes.  Science frames were first
sky-subtracted using a sky frame created by median-combining all of the
science data taken on a given night and then flat-fielded using a
normalized dome flat.  Individual frames were shifted and stacked to
form the final combined image.  The reduced science frame was
astrometrically calibrated using the 2MASS Point Source catalogue.  We
used a basic, six parameter, least-squares, linear transformation to
obtain an astrometric solution for the CPAPIR image.  The field-of-view
of CPAPIR is $\sim$30$''$ so there were over 1600 2MASS reference stars
to calibrate CPAPIR to 2MASS.  We limited our reference stars to the 450
stars with 12 $<$ $J$ $<$ 15 as objects in this intermediate magnitude
range transformed with the smallest residuals from epoch to epoch. The
solution reference stars were required to transform with total absolute
residuals against 2MASS of $<$ 0.2 pixels (see Faherty et al. 2008 for
more details).  We measure a proper motion of $\mu_\alpha$=$+$1404 $\pm$
8 mas yr$^{-1}$ and $\mu_\alpha$=$-$51 $\pm$ 18 mas yr$^{-1}$ over the
9.1 yr baseline between the CPAPIR image and the discovery 2MASS image.
The errors were computed by combining the residuals of the astrometric
solution, which accounted for 0.16 pixels, with the plate scale of
CPAPIR ($\sim$1$''$ pixel$^{-1}$) under the given baseline.  The
positional uncertainty for \obj\ was also calculated by comparing the
residuals of transforming the X,Y position for our target over
consecutive dithered images. However this uncertainty was negligible
compared to the contribution from the astrometric solution.

\section{Analysis}

\subsection{Spectral Classification}

Determining the spectral type of \obj\ is not straightforward because an
L subdwarf spectral classification scheme has yet to be rigorously
defined in either the red optical or the near infrared due to the few
examples known.  The depth of the 1.4 H$_2$O band and the shape of the
$H$ band (see Figure \ref{fig:NIRComp}) suggests that \obj\ has a near
infrared spectral type intermediate between that of \sdtwo\ and \sdone,
perhaps sdL4 to sdL6.  In order to derive a spectral type based on the
red optical spectrum, we followed \citet{2007ApJ...657..494B} and
compared the GMOS spectrum of \obj\ to the spectra of field L dwarf
spectral standards \citep{1999ApJ...519..802K}.  The best match is the
L5 dwarf DENIS-P J1228.2$-$1547 resulting in an optical spectral type of
sdL5 for \obj.  Note that a similar comparison in the near infrared is
impossible because L dwarf spectral standards have yet to be defined at
these wavelengths (see however Kirkpatrick et al., in preparation).
Figure \ref{fig:OptComp} shows the red optical spectra of both \obj\ and
DENIS -P J1228.2$-$1547 (grey).  The strengths of the CrH and FeH band
heads, width of the \ion{K}{1} doublet, and overall 8000$-$9000 \AA\
slope of the two spectra agree well.  Also shown in Figure
\ref{fig:OptComp} are the spectra of \sdone\ and \sdtwo\ along with
their respective L dwarf standards.  The strength of the features in the
spectrum of \obj\ are intermediate between that of \sdtwo\ and \sdone\,
in good agreement with its derived spectral type of sdL5.

\subsection{Kinematics}

\begin{figure} 
\includegraphics[width=3in]{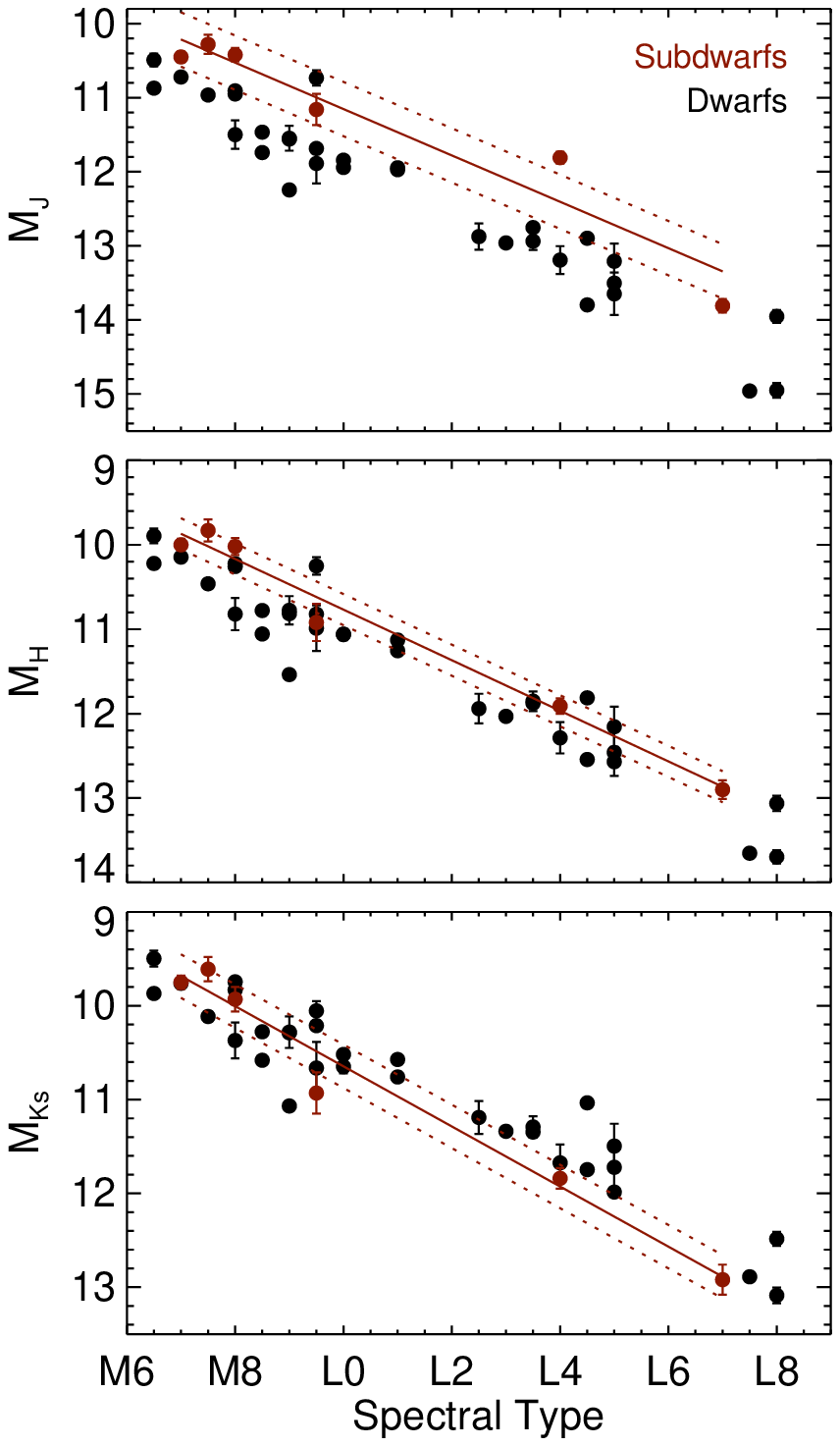}
\caption{\label{fig:AbsMags}The 2MASS $M_J$, $M_H$, and $M_{Ks}$ values
  for LHS 377 (sdM7), LSR 2036$+$5059 (sdM7.5), LSR J1425$+$7102 (sdM8),
  SSSPM 1013$-$1356 (sdM9.5), \sdtwo\ (sdL4) and \sdone\ (sdL7) are
  shown in red.  The linear fit to these values and the $\pm$1 $\sigma$
  range are shown as solid and dotted lines.  Shown in black are a
  sample of field dwarfs from \citet{2002AJ....124.1170D} and references
  therein that have not been identified as binaries in high resolution
  imaging.  All spectral types are based on optical spectroscopy.}
\end{figure}

Although the high proper motion and large radial velocity of \obj\
suggest that \obj\ is a high velocity star and thus possibly a member of
the halo population, a distance is required in order to compute space
velocities.  We estimated a spectrophotometric distance to \obj\ in the
following way.  First we performed a weighted linear fit of the 2MASS
$M_J$, $M_H$, and $M_{Ks}$ values with respect to spectral type of the
ultracool subdwarfs with known parallaxes
\citep{1992AJ....103..638M,2008arXiv0806.2336D,2008arXiv0811.4136S,2008ApJ...672.1159B},
LHS 377 (sdM7), LSR 2036$+$5059 (sdM7.5), LSR J1425$+$7102 (sdM8), SSSPM
1013$-$1356 (sdM9.5), \sdtwo\ (sdL4) and \sdone\ (sdL7) and find,

\begin{equation}
M_{J} = 8.02 + 0.313\times \mathrm{SpType}, \,\,\,\,\,\,  Cov = \left ( \begin{array}{cc} 0.137 & -0.0114  \\ -0.0114 & 0.00106 \\ \end{array} \right )
\end{equation}

\begin{equation}
M_{H} = 7.77 + 0.300\times \mathrm{SpType}, \,\,\,\,\,\,  Cov = \left ( \begin{array}{cc} 0.148 & -0.0125  \\ -0.0125 & 0.00120 \\ \end{array} \right )
\end{equation}

\begin{equation}
M_{Ks} = 7.44 + 0.320\times \mathrm{SpType}, \,\,\,\,\,\,  Cov = \left ( \begin{array}{cc} 0.172 & -0.0152  \\ -0.0152 & 0.00151 \\ \end{array} \right ),
\end{equation}
\noindent
where SpType=7 for M7, SpType=10 for L0, etc., and $Cov$ is the
covariance matrix of the fit.  For those dwarfs with two parallax
measurements, we used the weighted average of the two values to compute
their absolute magnitudes.  Figure \ref{fig:AbsMags} shows the absolute
magnitudes of the subdwarfs along with those of single field M and L
dwarfs from \citet{2002AJ....124.1170D} and references therein.  In
comparison to the field dwarfs, the L subdwarfs appear overluminous in
the $J$ band and underluminous in the $K$ band.  If we assume that the
effective tempereature scale of the L dwarfs and subdwarfs are similar
\citep[see however][]{2006ApJ...645.1485B,burgasser09a}, we can ascribe
this behavior to a change in metallicity.  At a fixed \teff\ and $g$, L
dwarfs become brighter in the $J$ band and fainter in the $K$ band with
decreasing metallicity \citep{2006ApJ...640.1063B} due primarily to the
increasing importance of CIA H$_2$ absorption in the $K$ band.

We estimate a distance of 57$\pm$9 pc for \obj\ using the absolute $J$
and $H$ magnitudes derived using Equations 1 and 2 for a spectral type
of sdL5 (\obj\ has only an upper limit in the $K_s$ band).  The value is
the weighted average of the two spectrophotometric distance estimates
and the error includes the covariance terms and an error of $\pm$0.5 a
subtype in the spectral type.  The estimated distance and proper motion
of \obj\ imply a tangential velocity $V_\mathrm{tan}$ of 379$\pm$60 km
s$^{-1}$ which is roughly an order of magnitude greater than the median
value of field L dwarfs \citep{2004AJ....127.2948V,2008arXiv0809.3008F}
and confirms that \obj\ is a member of the halo population.  Indeed
\citet{2008arXiv0809.3008F} found no MLT dwarfs with $V_\mathrm{tan}$
$>$ 170 km s$^{-1}$ in their sample of 634 L and T dwarfs and 456 M
dwarfs.  We have also computed $(u, v, w)$\footnote{We follow the
  convention that lower case letters refer to velocities with respect to
  the LSR while upper case letters refer to velocities with respect to
  the Sun \citep{1981gask.book.....M}.}  velocities of \obj\ with
respect to the Local Standard of Rest (LSR) following
\citet{1987AJ.....93..864J} updated with the J2000.0 galactic coordinate
transformations of \citet{1989A&A...218..325M}.  Following the
convention that the $u$ is positive towards the Galactic Center ($l$=0,
$b$=0), we find $(u, v, w)$=(94, $-$573, 125)$\pm$(10, 31, 53) km
s$^{-1}$ after correcting for a solar velocity with respect to the LSR
of (10.00, 5.25, 7.17) km s$^{-1}$ \citep{1998MNRAS.298..387D}.  The LSR
moves at $\Theta_\mathrm{LSR}$=220 km s$^{-1}$ \citep[$\Theta$ and $v$
point in the direction of Galactic rotation, $l$=90, $b$=0;
][]{1986MNRAS.221.1023K} in the rest frame of the Galaxy so \obj\ has a
highly retrograde orbit with $\Theta$=$-$353 km s$^{-1}$.  The stars in
the inner halo of the Galaxy have slightly prograde orbits with
$\Theta$=0 to 50 km s$^{-1}$ while the stars in the outer halo have
retrograde orbits with $\Theta$=$-$40 to $-$70 km s$^{-1}$
\citep{2007Natur.450.1020C}.  The rotation of \obj\ around the center of
the Galaxy clearly indicates that \obj\ is a member of the outer halo.

Based on its ($u$, $v$, $w$) velocities we also computed the Galactic
orbit of \obj\ following \citet{burgasser09a}.  We examined two
simplified versions of the Galactic potential model described in
\citet{1998MNRAS.298..387D} composed of spherically-symmetric halo and
bulge mass distributions and an axisymmetric, thin exponential disk.
The two Galactic mass models were parameterized according to
\citet[][Table 2.3]{2008gady.book.....B}, which fit the measured
rotation curve of the Galaxy but bracket the allowable ratios of disk to
halo mass at the Solar radius.  The orbit of \obj\ was integrated using
a second-order leapfrog method (kick-drift-kick) with a constant
timestep of 1 kyr over $\pm$1 Gyr centered on the present epoch.  Energy
was conserved to better than 1 part in $10^{-4}$ over the full length of
the simulation, with the error dominated by the resolution of the grid
on which the disk force and potential were interpolated.  The $Z$
component of angular momentum was conserved to 1 part in $10^{-13}$.

\begin{figure*} 
\centerline{\includegraphics[width=6in]{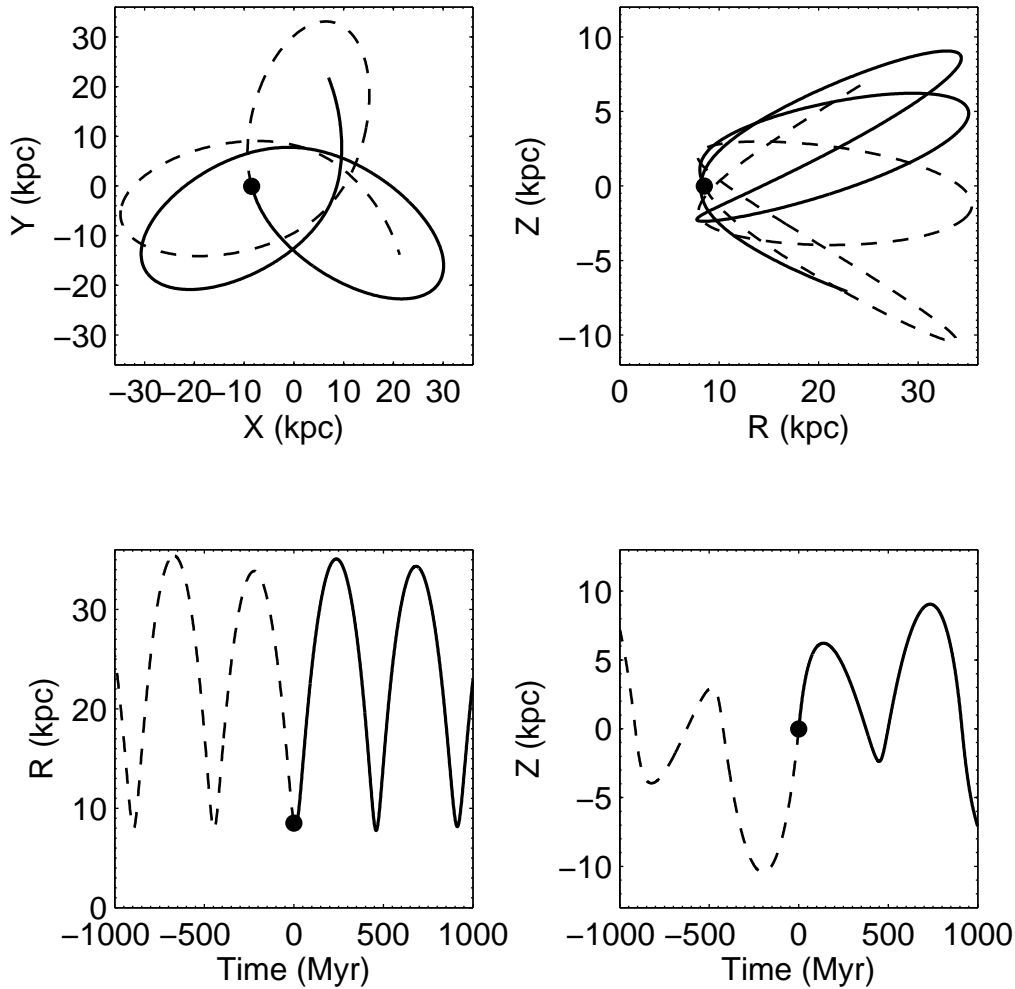}}
\caption{The Galactic orbit of \obj\ over $\pm$1~Gyr, based on the
  halo-dominant mass model of \citet[Model I;][]{2008gady.book.....B}.
  The upper panels show the orbit in ($X$,$Y$) and ($R$,$Z$)
  coordinates. The $X$ axis is along the line connecting the Galactic
  Center and the Sun while the $Z$ axis denotes height above the
  Galactic plane. The lower panels show the time evolution of $R$ and
  $Z$, with past motion denoted by dashed lines and future motion
  denoted by solid lines.  The position of the Sun is denoted with a
  solid circle.}
\end{figure*}

Figure 6 shows the orbit of 2MASS J0616-6407 based on the halo-dominant
mass model, plotted in the rest frame of the Galaxy.  The X coordinate
is defined to be positive towards the Galactic Center to align with our
definition of $u$, and the Sun is located at ($-$8.5, 0, 0.027) kpc
\citep{1986MNRAS.221.1023K,2001ApJ...553..184C}. Prograde motion in this
reference frame is counter-clockwise, so the orbit of \obj\ is clearly
retrograde, as well as wide and eccentric, with R extending from 8 to 36
kpc ($e$ $\sim$ 0.7 from max/min R) over the simulaiton period.  The
orbit of \obj\ also exhibits a fairly broad range of inclinations, with
Z spanning $\pm$10 kpc (max $|i|$ $\sim$ 15 degrees).  For the
disk-dominant mass model, the orbit is even wider (extending out to 40
kpc) but with similar eccentricity and inclination range.  The breadth
of this orbit and its retrograde motion support the conclusion that
\obj\ is a member of the outer halo, as this population dominates at R
$>$ 14-20 kpc \citep{2007Natur.450.1020C}. We note that our conclusions
remain unchanged if 2MASS J0616- 6407 is an equal magnitude binary.


\section{Discussion}

The Galactic orbit of \obj\ is distinctly different from that of other
late-type subdwarfs (e.g., LSR J0822$+$1700, \sdone, LSR J1425$+$7102,
\sdthree) which all have perigalacticons of $\lesssim$1 kpc and
apogalacticons near the Sun
\citep{2004ApJ...602L.125L,2008ApJ...672.1159B,2008ApJ...686..548D,burgasser09a}.
As noted by \citet{burgasser09a}, these other subdwarfs have kinematic
properties which are consistent with membership in the inner halo of the
Galaxy.  \obj\ is therefore the first late-type subdwarf that is a
member of the outer halo and therefore may represent a new class of L
subdwarfs.  In particular, the inner halo is comprised of stars with a
peak metallicity of [Fe/H] $\sim -$1.6 while the outer halo has a peak
near $-$2.2 \citep{2007Natur.450.1020C}.  This suggests that \obj\ may
be even more metal-poor than the other late-type subdwarfs known.
However, the techniques needed to derive accurate metallicities of
late-type subdwarfs are in their infancy \citep[e.g.,
][]{2007ApJ...657..494B,burgasser09a} and consequently an accurate
measurement of the metallicity of \obj\ must await future work.  Finally
we note that the discovery of additional outer halo ultracool subdwarfs
will be difficult given their intrinsic faintness and the large amount
of time that they spend away from the solar neighborhood.

\acknowledgements

We thank Bill Vacca for fruitful discussions about the reduction of
optical spectra and John Bochanski and Andrew West for clarifications on
the Galactic orbit geometry.  This publication makes use of data from
the Two Micron All Sky Survey, which is a joint project of the
University of Massachusetts and the Infrared Processing and Analysis
Center, and funded by the National Aeronautics and Space Administration
and the National Science Foundation, the SIMBAD database, operated at
CDS, Strasbourg, France, NASA's Astrophysics Data System Bibliographic
Services, the M, L, and T dwarf compendium housed at DwarfArchives.org
and maintained by Chris Gelino, Davy Kirkpatrick, and Adam Burgasser,
and the NASA/ IPAC Infrared Science Archive, which is operated by the
Jet Propulsion Laboratory, California Institute of Technology, under
contract with the National Aeronautics and Space Administration.  OSIRIS
is a collaborative project between the Ohio State University and Cerro
Tololo Inter-American Observatory (CTIO) and was developed through NSF
grants AST 90-16112 and AST 92-18449. CTIO is part of the National
Optical Astronomy Observatory (NOAO), based in La Serena, Chile. NOAO is
operated by the Association of Universities for Research in Astronomy
(AURA), Inc. under cooperative agreement with the National Science
Foundation.

{\it Facilities:} \facility{Gemini (GMOS)}, \facility{SOAR (OSIRIS)},
\facility{LDSS-3 (Magellan)}, \facility{CPAPIR (CTIO)}.

\clearpage

\bibliographystyle{apj}
\bibliography{ref,tmp}

\end{document}